# Interactions between Magnetic Nanowires and Living Cells : Uptake, Toxicity and Degradation

**Malak Safi[1], Minhao Yan[1], Marie-Alice Guedeau-Boudeville[1], Hélène Conjeaud[1], Virginie Garnier-Thibaud[2], Nicole Boggetto[3], Armelle Baeza-Squiban[4], Florence Niedergang[5,6,7], Dietrich Averbeck[8] and Jean-François Berret[1]**

[1] Matière et Systèmes Complexes, UMR 7057 CNRS Université Denis Diderot Paris-VII, Bâtiment Condorcet10 rue Alice Domon et Léonie Duquet, 75205 Paris (France)
[2] Service de Microscopie Electronique, Institut de Biologie Intégrative, IFR 83 Université P. et M. Curie, 9 quai St Bernard 75252 Paris cedex
[3] Université Paris Diderot-Paris 7, ImagoSeine Bioimaging Core Facility, Jacques Monod Institute, 75013 Paris, France.
[4] Université Paris Diderot-Paris 7, Unit of Functional and Adaptive Biology (BFA) CNRS EAC 4413, Laboratory of Molecular and Cellular Responses to Xenobiotics, Bâtiment Buffon, 5 rue Thomas Mann, 75013 Paris, France
[5] Inserm, U1016, Institut Cochin, Paris, France.
[6] CNRS, UMR 8104, Paris, France.
[7] Univ Paris Descartes, Paris, France
[8] Institut Curie-Section de Recherche, Centre Universitaire Paris-Sud, Bâtiment 110, 91405 Orsay, France

E-mail: jean-francois.berret@univ-paris-diderot.fr

**Abstract :** We report on the uptake, toxicity and degradation of magnetic nanowires by NIH/3T3 mouse fibroblasts. Magnetic nanowires of diameters 200 nm and lengths comprised between 1 µm and 40 µm are fabricated by controlled assembly of iron oxide ($\gamma$-Fe$_2$O$_3$) nanoparticles. Using optical and electron microscopy, we show that after 24 h incubation the wires are internalized by the cells and located either in membrane-bound compartments or dispersed in the cytosol. Using fluorescence microscopy, the membrane-bound compartments were identified as late endosomal/lysosomal endosomes labeled with lysosomal associated membrane protein (Lamp1). Toxicity assays evaluating the mitochondrial activity, cell proliferation and production of reactive oxygen species show that the wires do not display acute short-term (< 100 h) toxicity towards the cells. Interestingly, the cells are able to degrade the wires and to transform them into smaller aggregates, even in short time periods (days). This degradation is likely to occur as a consequence of the internal structure of the wires, which is that of a non-covalently bound aggregate. We anticipate that this degradation should prevent long-term asbestos-like toxicity effects related to high aspect ratio morphologies and that these wires represent a promising class of nanomaterials for cell manipulation and microrheology.

**Keywords** : magnetic nanoparticles, nanowires, assembled nanostructures, cytotoxicity, internalization

Inorganic nanomaterials and particles with enhanced optical, mechanical or magnetic attributes are currently being developed for a wide range of applications, including catalysis, photovoltaics, coating and nanomedicine. In nanomedicine, iron oxide or semiconductor nanocrystals are nowadays used as contrast agents for imaging or as drug delivery vectors.[1] More generally, nanomaterials of different shapes and sizes are looked upon as promising tools for targeting, diagnostic and therapy at cell scale. In this context, the interactions of nanomaterials with living organisms are investigated extensively. It is believed that submicron size objects, both organic or inorganic can induce reactive oxygen species (ROS) which are at the origin of various pathological disorders, including cardiovascular and neurodegenerative diseases. As shown in several reviews,[2,3] the effects of nanomaterials on living cells and tissues and possible health risks have not yet been fully evaluated.

Among the wide variety of nanomaterials designed so far, magnetic nanowires have received considerable attention because of their importance in cell manipulation, microfluidics and micromechanics.[4-10] Nanowires are anisotropic colloidal objects with submicronic diameters and lengths in the range of 1 to 100 μm. In specific applications such as cell separation, ferromagnetic nickel (Ni) nanowires were shown to outperform magnetic beads of comparable volume.[6] An efficient strategy for the synthesis of nanowires consists in the electrodeposition of nickel or iron atoms into thin alumina-based porous templates with cylindrical holes.[11] At dissolution of the template, rigid ferromagnetic nanowires are produced and dispersed in water-based solvents. These wires were used in cell guidance,[4] cell separation[12] and microrheology experiments.[8] One major drawback encountered with nickel and iron nanowires is that these objects carry a permanent magnetic moment and are thus susceptible to aggregate in solution because of magnetic dipolar attraction.[4,5]

Concerning the interactions with living cells, Hultgren *et al.*[6] have shown that nickel nanowires were easily internalized by NIH/3T3 cells and suggested the integrin-mediated phagocytosis pathway as portal of entry.[13] The uptake was optimized when the length of the nanowires was matched to the diameter of the cells in culture. Transmission electron microscopy showed that in the cytoplasm, the nanowires were not surrounded by a lipid bilayer envelope, indicating that the nanowires were trafficked into the cytoplasm. These results were in good agreement with those of Champion *et al.*[14] who examined the role of target geometry on the phagocytosis and observed that the local curvature at the point of initial contact dictated whether cells initiated phagocytosis or not. Using the same wire/cell model as in Ref.[6], Fung *et al.* have found that the internalized nickel nanowires could induce cell death by magnetic actuation.[15] Magnetic torques applied on wires located inside the cells were indeed sufficient to initiate propeller-like rotations and provoke mechanical mixing of the intracellular medium. More recently, Song *et al.*[5] explored the cytotoxicity and cellular uptake of iron nanowires on HeLa cancer cells. These authors suggested two main mechanisms for the uptake. For lengths below a few microns, the wires were engulfed by non-specific pinocytosis and remained in endosomes.[16,17] Beyond, the wires were found to be dispersed in the cytoplasm, suggesting an entry mechanism based on the perforation of the cellular membrane. For sake of completeness, it should be mentioned that the effects of particle shape on the internalization pathways were also reported for particles with lower anisotropy ratio.[18-20]

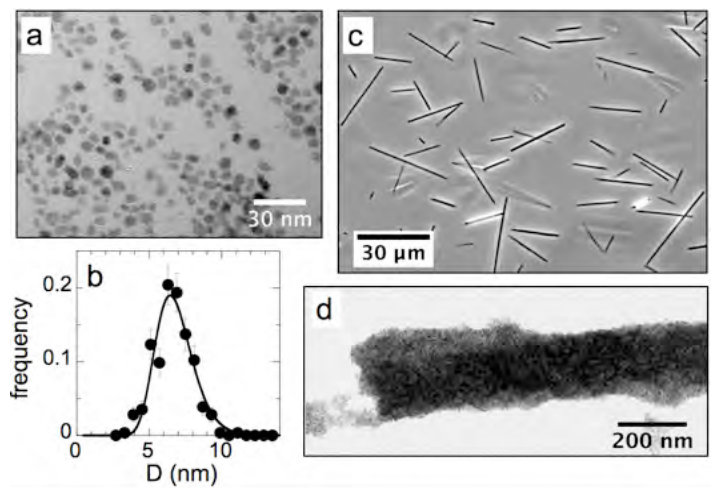

**Figure 1**: a) Transmission electron microscopy of iron oxide ($\gamma$-$Fe_2O_3$, maghemite) at the magnification of ×120000. b) Size distribution of the $\gamma$-$Fe_2O_3$ fitted by a log-normal distribution function with median diameter $D_{NP}$ = 6.7 nm and polydispersity $s_{NP}$ = 0.2. c) Phase-contrast optical microscopy images (40×) of a dispersion of nanostructured wires made from 6.7 nm $\gamma$-$Fe_2O_3$ particles in the absence of magnetic field. d) TEM image of the extremity of a wire showing the individual particles tightly held together and forming a core cylindrical structure (26000×).

In the present paper, we exploit a simple and versatile waterborne synthesis process to generate magnetic nanowires.[21, 22] Highly persistent wires of diameters 200 nm and lengths comprised between 1 µm and 40 µm were fabricated by controlling the assembly of sub-10 nm iron oxide nanoparticles (Fig. 1). These magnetic nanowires are different from the ferromagnetic electrodeposited wires described previously. The wires are superparamagnetic *i.e.* they do not carry a permanent magnetic moment, which prevents their spontaneous aggregation in a dispersion.[21, 23] Since these wires are aimed to be used as micromechanical tools at the cellular level, their interactions with living cells need to be investigated. Using NIH/3T3 mouse fibroblasts, we investigate these interactions and show that the wires are internalized by the cells, and found either in late endocytic compartments or dispersed in the cytosol. Extensive toxicity assays testing the mitochondrial activity, cell proliferation and production of reactive oxygen species reveal that the wires do not display acute toxicity towards the cells in short-term. In this study are also compared the effects of the wires on the mouse fibroblasts and those of their particulate constituents.

# Results and discussion
## Nanowires are internalized by the cells

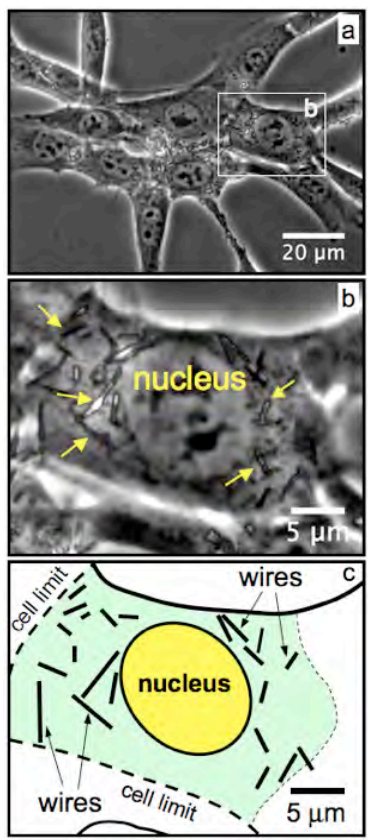

**Figure 2 :** *a) Phase contrast microscopy images of NIH/3T3 fibroblasts cells treated with 15 μm nanowires for 24 h at a concentration of 30 wires/cell. b) A close-up of the area delimited by a rectangle in a) emphasizes the perinuclear region of a unique cell. Wires of various lengths and orientations are indicated by arrows. c) Schematic representation of the area in b) showing the presence of 22 clearly visible wires with size between 1 to 7 μm.*

Fig. 2a shows a cluster of cells that were incubated with magnetic nanowires of length 15 μm and at 30 wires per cell (Table I). In this work, the nanowire concentration was defined by the ratio of the number of wires incubated by the number of cells plated, and by the iron molar concentration [Fe]. This latter definition allows comparison with the $PAA_{2K}-\gamma-Fe_2O_3$ data or with data from the literature.[24-27] In the present case, 30 wires per cell correspond to [Fe] = 0.5 mM. Because of the wire polydispersity, the fibroblasts were actually exposed to threads comprised between 1 and 40 μm. The cells were found to maintain their morphology and adherence properties after a 24 h exposure. Fig. 2b displays a close-up of the area delimited by a rectangle in Fig. 2a and emphasizes the perinuclear region of a single cell. There, elongated threads with various lengths and orientations are observed (arrows). A schematic representation of this area is provided in the lower panel of Fig. 2, showing the presence of 22 clearly visible wires with size between 1 to 7 μm (Fig. 2c). The location of the wires in the cytoplasm and not outside in the supernatant was confirmed in two ways. The Brownian motions of wires were recorded as a function of the time and compared to those of wires dispersed in water. It was found that the rotational diffusion constant was much slower for the internalized objects (approximately by a factor of 200, see Supporting Information SI-1), in good agreement with recent estimations of the viscosity of the intracellular matrix.[28]

Furthermore, we applied an external magnetic field to the sample and could observe wire reorientations that were constrained as compared to their motions in water, indicating again that the wires were inside the cytosol (see Movie#1 and 2 in SI).[8]

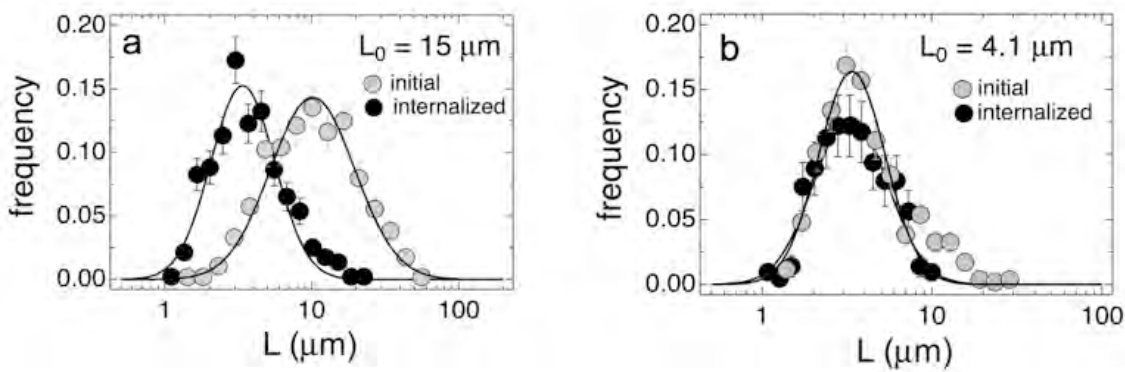

*Figure 3 :* Length distributions of internalized wires as compared to their initial forms. The distributions were found to be log-normal with median length and polydispersity noted $L_{wire}$ and s, respectively. a) Initial values : $L_{wire}$ = 15 µm, s = 0.65; internalized values : $L_{wire}$ = 4.3 µm, s = 0.5. b) For $L_{wire}$ = 4.1 µm and s = 0.45, initial and internalized distributions were identical.

| nanowires | exp. conditions | $L_{wire}$ µm | s | techniques used |
|---|---|---|---|---|
| 15 µm nanowires | initial | 15.6±1 | 0.65±0.5 | OM |
| 15 µm nanowires | internalized | 4.3±0.2 | 0.50±0.5 | OM |
| 15 µm nanowires | internalized | 1.3±0.2 | 0.45±0.5 | TEM |
| 4 µm nanowires | initial | 4.1±0.1 | 0.45±0.5 | OM |
| 4 µm nanowires | internalized | 4.2±0.2 | 0.50±0.5 | OM |

*Table I :* Characteristic length ($L_{wire}$) and polydispersity (s) for the batches used in this study. For internalized wires, the experimental conditions were an incubation time of 24 h and a number of wires per fibroblast of 30. The length distributions were measured from phase-contrast optical microscopy (OM) and transmission electron microscopy (TEM) data. With TEM, the wires appeared shorter than their internalized length (1.3 µm versus 4.3 µm) and exhibited sharp and diffuse extremities (Fig. 9). These findings were attributed to the fact that the wires not in the plane of the cut were shortened by the sample preparation.

We then evaluated the length distribution of the internalized wires and compared it to its initial form (Figs. 3). Each distributions was determined from a panel of more than 250 objects. The initial distribution in Fig. 3a was found to be log-normal with a median length of $L_{wire}$ = 15.6 ± 0.2 µm and a polydispersity s = 0.65 ± 0.05. By contrast, the internalized wires were found to be much smaller, with a median value $L_{wire}$ = 4.3 ± 0.2 µm and a narrower dispersity s = 0.50 ± 0.05 (table I). When exposed to polydisperse wires, the cells operated a sorting process as a function of the length. In a second experiment, wires with a smaller size distribution ($L_{wire}$ = 4.1 µm and s = 0.45) were incubated with the cells. In this case, the length distributions of the initial and internalized wires matched precisely, indicating here that the sorting process did not take place (Fig. 3b and Table I). These results suggest that the internalization of nanowires was optimized when their length corresponded to the average size of the adherent cells. Further evidences of internalization and size sorting were provided by TEM and immunofluorescence experiments.

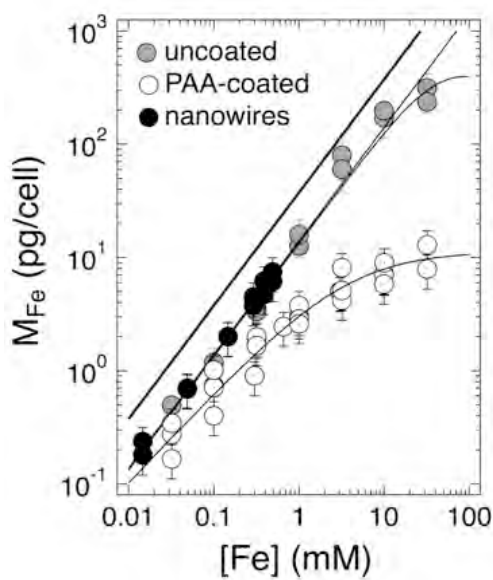

*Figure 4 :* Amount of internalized or adsorbed iron oxide $M_{Fe}$ (pg/cell) determined following the MILC protocol (see text for details). The upper thick straight line depicts the maximum amount of iron that can be uptaken by the cells for a molar concentration [Fe] in the supernatant. Empty circles : $PAA_{2K}$–$\gamma$-$Fe_2O_3$; grey circles : uncoated $\gamma$-$Fe_2O_3$; closed circles : 4 μm nanowires.

## Quantitative determination of internalized iron oxide

The amount of internalized or adsorbed iron oxide was determined following the protocol MILC (**M**ass of metal **I**nternalized/Adsorbed by **L**iving **C**ells) which makes use of UV spectrophotometry to calculate the iron concentration from pelleted cells dissolved in concentrated HCl (SI-2). The results were expressed in picograms of iron per cell. Fig. 4 compares the [Fe]-dependencies of internalized/adsorbed amounts $M_{Fe}$ for uncoated, $PAA_{2K}$ coated particles and 4 μm nanowires. For this batch, [Fe] = 0.5 mM corresponds to 120 wires per cell. The continuous lines through the data points are guides for the eyes. The thick straight line in Fig. 4 depicts the maximum amount of iron that could be taken up by the cells, *i.e.* the amount of iron added to the supernatant divided by the number of cells in the assay. For $3\times10^6$ fibroblasts exposed to a [Fe] = 1 mM solution, this maximum amount lies at 37 pg/cell. For the uncoated particles and for the nanowires, $M_{Fe}$ increases linearly with the iron concentration and represents about 35% of the maximum value. Optical microscopy carried out 24 h after incubation and thorough washing showed that none of the 4 μm wires were adsorbed at the cell membranes, leading to the conclusion that the $M_{Fe}$-data in Fig. 4 represent the amount of internalized iron under the form of nanowire. For uncoated $\gamma$-$Fe_2O_3$, the large quantities detected up to [Fe] = 10 mM were explained by the precipitation of the particles in the culture medium. This aggregation produced large and compact clusters in the micrometer range (1 – 20 μm)[29] that precipitated on the cells and were adsorbed on the plasma membrane. Thorough washing with PBS buffer did not desorb these aggregates. For $PAA_{2K}$–$\gamma$-$Fe_2O_3$, the mass of iron internalized remained at a lower level (10 pg/cell for [Fe] = 10 mM). Thanks to the polymer brushes tethered on their surfaces, the particles were found to be very stable in the physiological medium, preventing their precipitation and their adsorption onto the cells.[27]

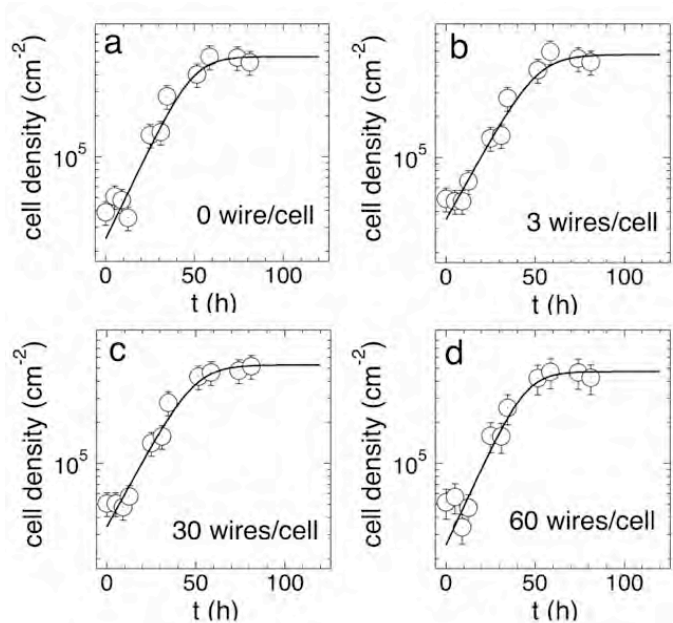

*Figure 5*: *Number density n(t) as a function of time for fibroblasts plated with 4 μm nanowires at increasing concentrations, corresponding to 0 (a), 3 (b), 30 (c), and 60(d) wires per cell. The continuous lines were obtained using a modified exponential growth model (Eq. 1), with adjusting parameters $n_0$ and $n_S$ which are the initial and final cell densities and $\tau_D$ the duplication time (Table II).*

Toxicity assays

Cytoxicity studies determine the cellular alterations or damages of vital functions induced by xenobiotics. In the present work, normal mitochondrial metabolic functions were probed by MTT assays and mitotic capacity by testing cell proliferation. As for the proliferation, the fibroblasts were plated with 4 μm nanowires in Petri dish cultures at increasing concentrations, corresponding to 0, 3, 30, and 60 wires per cell. NIH/3T3 cell densities noted n(t) were counted in a Malassez chamber at regular time periods from day 0 and to day 5. The data obtained are shown in Fig. 5a-d. Over the first 48 h, cell densities exhibited an exponential growth and then a saturation ($n_S$). The temporal dependencies of n(t) were adjusted by a modified exponential growth model[27] of the form :

$$n(t) = n_0 2^{t/\tau_D} \left(1 + \frac{n_0^2}{n_S^2}\left(2^{2t/\tau_D} - 1\right)\right)^{-1/2} \quad \text{(Eq. 1)}$$

where $n_0$ denotes the initial cell number and $\tau_D$ the duplication time. This model takes in to account the slowing-down of the growth as the surface coverage reached saturation. The continuous lines in Figs. 5 provided best-fit calculations using Eq. 1. The fitting parameters listed in Table II show that the cell duplication times ($\tau_D = 11 - 13$ h) and the final densities ($n_S = 5 \times 10^5$ cm$^{-2}$) were not affected by the presence of the wires, and that the cells proliferated normally over 80 h. Proliferation assays performed on cells incubated with the single PAA$_{2K}$–γ-Fe$_2$O$_3$ particles have demonstrated similar results (SI-3).

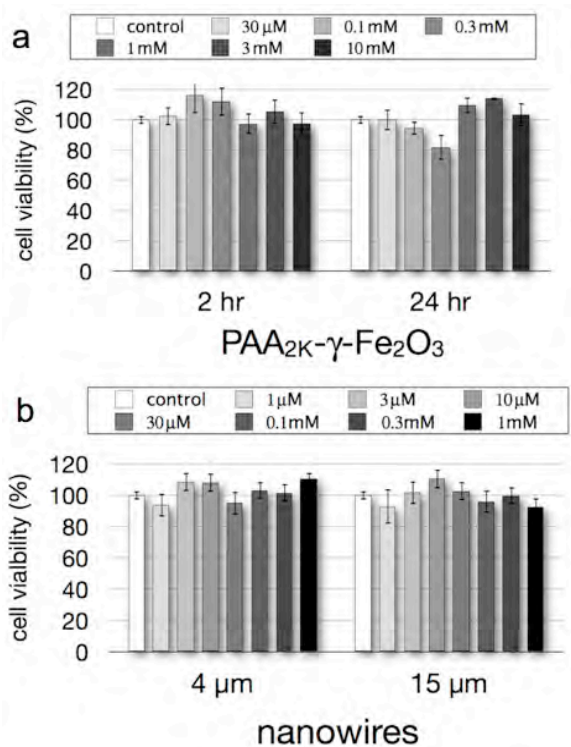

*Figure 6: MTT (3-(4,5-dimethylthiazol-2-yl)-2,5-diphenyl tetrazolium bromide) viability assays conducted on NIH/3T3 cells incubated with a) $PAA_{2K}$–γ-$Fe_2O_3$ during 2 h and 24 h (a) and b) with 4 µm and 15 µm nanowires during 24 h. in this work, the nanowire concentration was defined by the ratio of the number of wires incubated per cell. To allow comparison with data from the literature, it was also expressed in terms of iron molar concentration [Fe].*

MTT viability assays were conducted on NIH/3T3 cells for the $PAA_{2K}$–γ-$Fe_2O_3$ particles ([Fe] = 10 µM - 10 mM) and for the 4 and 15 µm nanowires ([Fe] = 0.1 µM - 1 mM). The comparison between this two types of nanomaterials aimed to identify the role of size and morphology on the cell survival. Exposure times were set at 2 h and 24 h for the particles and 24 h for the wires. As shown in Fig. 6 for both systems, the viability remained at a 100% level within the experimental accuracy. These findings indicate a normal mitochondrial activity for the cultures tested. The results on the single unprecipitated $PAA_{2K}$–γ-$Fe_2O_3$ are in good agreement with earlier reports.[25, 27, 30] For the nanowires, the data also confirm those obtained recently on parent micron-size materials, such as silica[19] or carbon nanotubes[31] and iron wires.[5] In conclusion, the proliferation and MTT assays show convincing results and ensure the suitability of the wires for biomedical and biophysical applications.

Oxidative stress

Reactive oxygen species (ROS) production was evaluated using the fluorescence changes of the permeant dye dihydroethidium (DHE) induced after oxidation by intracellular superoxide anions.[32] DHE exhibits blue-fluorescence in the cytosol until it is oxidized, while oxidized products intercalate within the DNA and exhibit bright red fluorescence. Cells were treated for 4 h at different wire concentrations (20, 70 and 170 wires per cell corresponding to [Fe] = 0.1, 0.3 and 0.7 mM). Negative and positive controls were performed using untreated cells and cells incubated with hydrogen peroxide ([$H_2O_2$] = 0 – 5 mM), respectively. Additional *ex vivo* control experiments showed that neither the particles nor the wires exhibited oxidative effects on the dye molecule used and did not modify its fluorescence (SI-4). Individual cell fluorescence and scatter properties were analyzed by flow cytometry. Quantification of the percentage of cells displaying red fluorescence (cells able to oxidize the probe due to intracellular ROS activity) as a function of the iron oxide concentration revealed that exposure to nanomaterials only marginally increased the number of responding cells (Fig. 7a). Both untreated and cells exposed to nanomaterials generated a low level of fluorescence, ranging from 1% to 1.8% of DHE positive cells. These values were inferior to those of positive control. $H_2O_2$ treated cells exhibited a strong concentration dependence of DHE positive cells, culminating at 80% for [$H_2O_2$] = 2 mM. Beyond 2 mM, the cells started to die and their numbers decreased (Fig. 7b). These results indicate that the number of NIH/3T3 fibroblasts that produce reactive oxygen species did not significantly increase upon 4 h treatment with wires or with $PAA_{2K}$–$\gamma$-$Fe_2O_3$ particles.

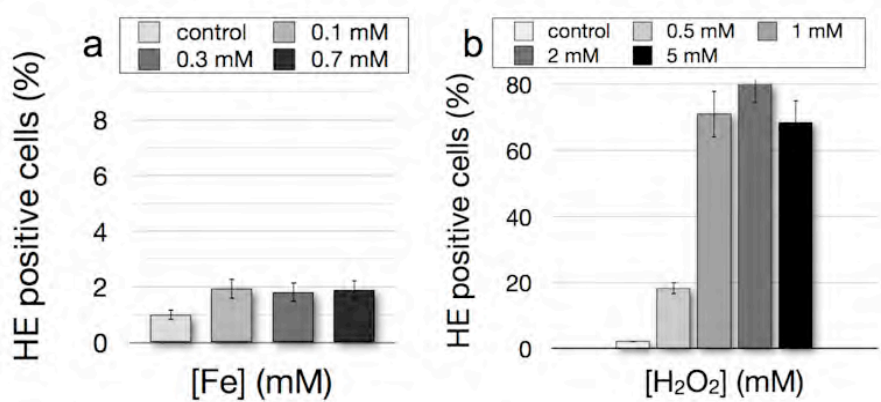

*Figure 7 : Flow cytometric quantification of intracellular ROS as a function of a) wire concentration and b) hydrogen peroxide concentration (control). The average wire length was $L_{wire}$ = 4.1 μm.*

Transmission electron microscopy

The fibroblasts seeded with the 15 μm nanowires were further investigated by TEM. Fig. 8 and 9 provide representative images obtained from treated fibroblasts. The experimental conditions were an incubation time of 24 h and a number of wires per cell of 30 ([Fe] = 0.5 mM). A careful analysis of the TEM data allowed us to classify the nanomaterials inside the fibroblasts under three categories :
1. Entire or pieces of nanowires,
2. Dense clusters of nanoparticles,
3. Single nanoparticles

Interestingly, all three states were found in intracellular compartments or directly in the cytosol. None of these three states were found in the nuclei. The most frequent configuration was that of nanowires directly dispersed in the cytosol. Fig. 8a and 8c provide instances of wires enclosed in intracellular compartments. In both cases, a lipidic membrane can be distinguished and form a barrier towards the cytosol. The sizes of the subcellular regions in Figs. 8 are large, 700 nm and 1 – 2 µm respectively. Close-up views of the delimited areas indicate that the wires have been degraded. Fig. 8b shows a wire that was bent so as to fit into the spherical compartment. In the left-hand region, the wire started to thin out and individual iron oxide nanoparticles were released in the surrounding fluid (red arrow in Fig. 8b and SI-6). Fig. 8d illustrates the case of a wire that was cut into shorter pieces, again to fit the vesicular dimensions. The loss of integrity of the wires could be explained by the decrease in pH occurring in lysosomal compartments. Stability assays of wires dispersed in aqueous solution as a function of the pH (pH 1.5 – 9) were conducted outside the cells and revealed no changes in wire morphology after 4 days (SI-5), indicating that other processes of degradation were taking place inside the cells.[33]

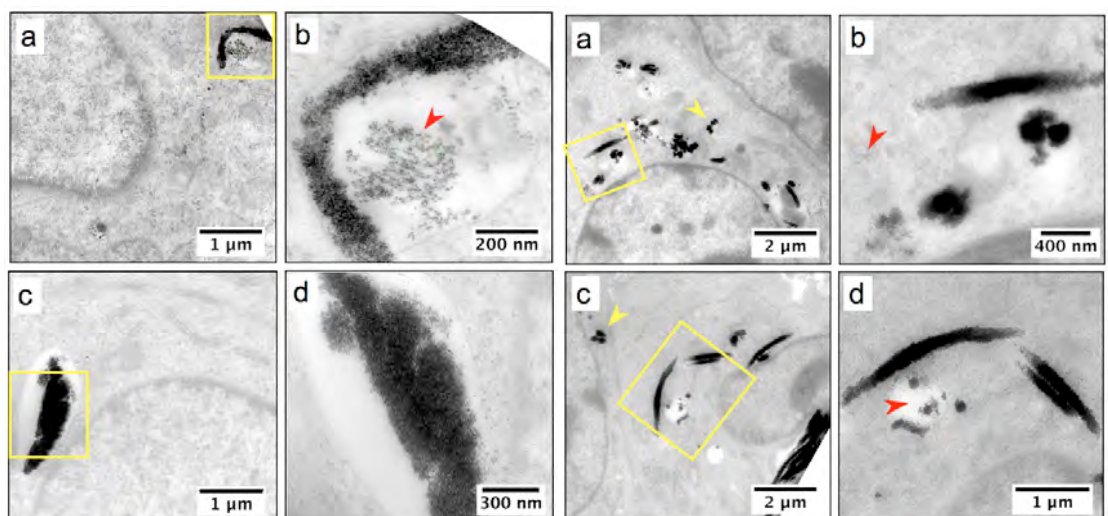

*Figure 8 :* Representative TEM images NIH/3T3 fibroblasts seeded during 24 h with 30 wires per cell ($L_{wire}$ = 15 µm). In these images, the wires are enclosed in intracellular compartments and shown at different length scales. The sizes of the membrane bound compartments are 700 nm (a) and 1 – 2 µm (c). The red arrow shows nanoparticles resulting form the degradation of the wirelike aggregates (see also SI-6).
*Figure 9 :* Representative TEM images NIH/3T3 fibroblasts in conditions similar to those of Fig. 8. In these pictures, the wires are dispersed into the cytosol and shown at different length scales. Dense aggregates (yellow arrow) and single nanoparticles (red arrow, see also SI-6) located either in endosomes or in the cytosol were also observed.

Among the data obtained by TEM, a frequent situation was that of wires or pieces of wires directly dispersed into the cytosol. Examples are highlighted in Fig. 9. In such cases, the nanostructures were not surrounded by any visible membrane. The wires also appeared to be shorter than their initial length, and exhibited sharp and diffuse extremities (Fig. 9b and 9d). A statistical analysis of the nanomaterials inside the NIH/3T3 revealed that the wires seen by TEM had an average length $L_{wire}$ = 1.3 µm (s = 0.45), that is lower than the size determined by optical microcopy (Fig. 3). This observation was attributed to the fact that the microtomed sections of cells were 90 nm-thick, and that wires not in the plane of the cut were shortened by the sample preparation.[34] In addition to the pieces of wires, aggregates of particles were also observed, either in compartments or located in the cytosol. The aggregates were spherical, dense and relatively monodisperse, with average diameter $D_{cluster}$ = 200 nm (yellow arrows in Fig. 9a and 9c). Fig. 10 compares the distribution of the initial 15 µm wires (Fig.

10a) to that of the wires found inside the cells (Fig. 10b) and to that of the clusters (Fig. 10c). It is important to recall at this point that the nanowire dispersions used for the seeding were thoroughly washed before use and contained neither submicronic aggregates nor single unassociated nanoparticles. Fig. 10 demonstrates that after a 24 h incubation, the initial wire population was split into two sub-populations, one of wires and one of aggregates.

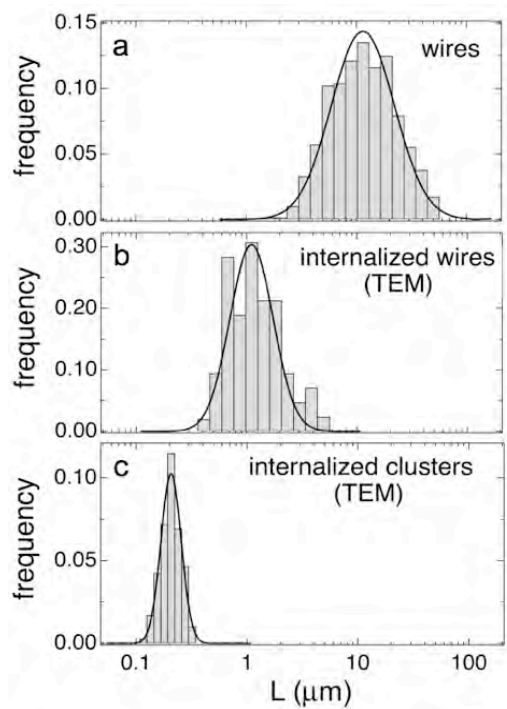

***Figure 10 :*** *a) Initial length distribution of nanowires used to incubate the fibroblasts ($L_{wire}$ = 15.6 µm and s = 0.65). b) Length distribution of wires or pieces of wires observed by TEM inside the cells ($L_{wire}$ = 1.3 µm and s = 0.45). c) size distribution of the nanoparticle clusters found by TEM inside the cells. Their average diameter was 200 nm. The proportions of the wire and cluster populations was estimated in a ratio 2:1.*

The proportions of each was estimated in a ratio 2:1. One-third of all iron oxide detected by TEM was thus under the form of nanoparticle clusters. This large proportion indicates that the clusters seen cannot be wires cut perpendicular to their long axis. At smaller amounts, single and isolated nanoparticles were also found in the cells, in vesicles and in the cytosol (red arrows in Fig. 9b and 9d, and SI-6). Put together, these findings lead to the conclusion that the wires were degraded by the cells. Similar results were obtained recently on carbon nanotubes which degradation was stimulated *in vitro* by myeloperoxidase enzymes. With carbon nanotubes, the biodegraded materials did not generate any inflammatory pulmonary response in mice, as compared to the pristine ones.[33] Note that in the present study, the fibroblasts did not need to be chemically stimulated, as in ref.[33], and that the degradation process was rather fast. It was nevertheless slow enough to allow the wires to be manipulated by an external field, and to allow cell separation or microrheology experiments on living cells.

The exposure of cells to $PAA_{2K}$–$\gamma$-$Fe_2O_3$ single nanoparticles was also investigated with TEM for comparison. Figs. 11a and 11b depict the representative behavior of particles at different length scales inside the NIH/3T3 fibroblasts. In marked contrast to the wires, the particles were exclusively found in membrane bound compartments of average size 500 nm. Note in Fig. 11b that the particles were randomly spread inside the endosome and not aggregated, as with cells treated with nanowires. Single particles in the cytosol or in the nucleus were not found. The differences between the internalized nanoparticles and nanowires suggest distinct mechanisms of entry into the cells or different fates of the intracellular compartments.

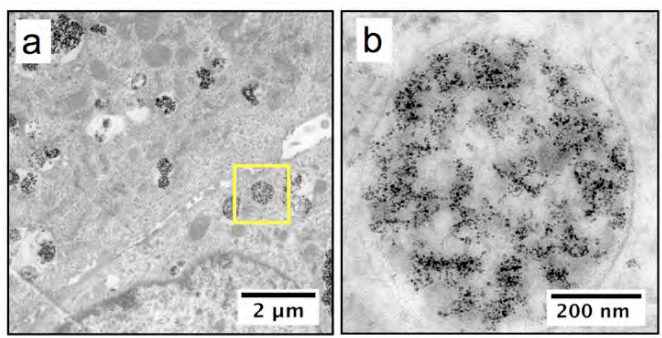

**Figure 11 :** a) TEM image NIH/3T3 fibroblasts incubated with $PAA_{2K}$–$\gamma$-$Fe_2O_3$ nanoparticles at [Fe] = 10 mM during 24 h. In these assays, the internalized particles were found only in membrane-bound compartments of average size 520 nm. The results of a statistical study performed on the size of the endosomes are shown in SI-10. Single particles in the cytosol were not found. b) A close-up of the area delimited by a square in (a).

## Immunofluorescence and localisation

To further analyze the subcellular localization of the nanowires in cells, NIH/3T3 cells were incubated with the magnetic nanowires, at the ratio of 30 wires per cell for 24h, then fixed and stained with antibodies to detect the Lysosomal Associated Membrane Protein (Lamp1), a marker of late endosomal/lysosomal endosomes. The cells were examined under an inverse motorized microscope providing Z-stacks of images taken every 0.2 μm. Fig. 12 displays two different regions comprising several fibroblasts. The observations were made in phase contrast (Figs. 12a and 12c) and in fluorescence (Figs. 12b and 12d). The corresponding Z-stacks can be visualized in the Supporting Information. In phase contrast, Fig. 12 shows a large number of nanowires of sizes 1 – 10 μm located in the cytoplasm. None of them were found in the nuclei. In contrast to Fig. 2 however, some wires are bent and exhibit kinks along their length. The modifications of the wire structure was attributed to the fixation and labeling protocols that were followed for the preparation of the cells. Figs. 12b and 12d display fluorescent images of the same fields, emphasizing the presence of late endosomes/lysosomes within the cytoplasm. The majority of these LAMP1-positive compartments are of spherical shape, except for a few of them which are elongated. The right panels in Fig. 12 present ranges of interest ($a_i$, $b_i$, $c_i$ and $d_i$ with i = 1,2) illustrating the localization of the nanowires in the anisotropic Lamp-1-positive fluorescent compartments. The blue arrowheads indicate wires that are detected by fluorescence and phase contrast images, whereas the red arrowheads show wires that are only seen in phase contrast. The elongated LAMP1-positive compartments have lengths between 1 and 8 μm. The proportion of LAMP1-positive elongated compartments with respect to the total number of internalized nanowires was estimated at 14 ± 5 %. These results are in excellent agreement with the TEM data, and suggest that only a fraction of nanowires are in Lamp-1-positive membrane-bound compartments 24h after internalization. 3D reconstructions of the close-up sections b1 and d2 obtained using the IsoSurface function of Imaris 5.7 software (Bitplane AG) are available in Supporting Information (SI-7).

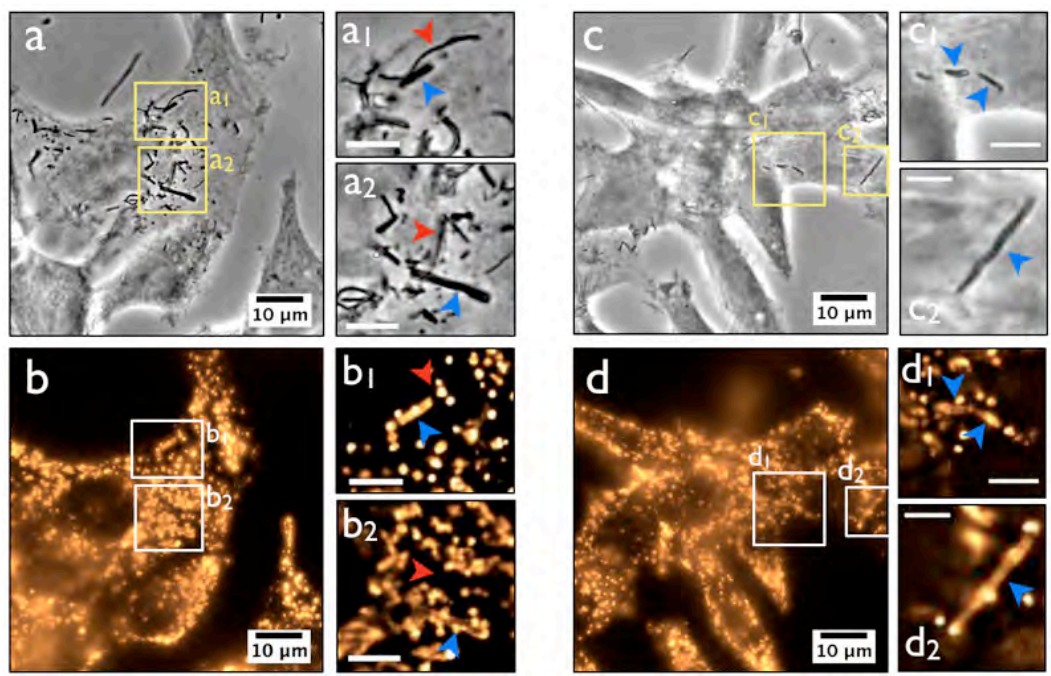

***Figure 12*** *: Cells were incubated with 30 nanowires per cell for 24 h at 37° C, then fixed, permeabilized and labeled with anti-Lamp1 antibodies, followed by Cy3-F(ab')$_2$ anti-rat IgG antibodies. Nanowires were detected by phase constrast (a, c). Z stacks of wide-field fluorescent images were acquired using a piezo and one Z projection of maximum intensity is shown (b,d). Images were deconvolved and close-up views in one medial optical section are shown ($a_i$, $b_i$, $c_i$ and $d_i$ with i = 1,2). Blue (resp. red) arrowheads indicate nanowires that are (resp. not) surrounded by Lamp1 staining. The bars in the right panels are 5 µm except for $c_2$ and $d_2$ (3 µm). 3D reconstructions obtained using the IsoSurface function of Imaris 5.7 software (Bitplane AG) are illustrated in SI-7.*

# Conclusion

In this paper, we evaluated the interactions and toxicity of magnetic nanowires with respect to NIH/3T3 mouse fibroblasts. Magnetic nanowires are a particularly promising class of nanomaterials since they can be used in confined geometries for microrheology and microfluidics and in general for cell manipulation. The wires studied here were different from the nanowires classically generated by electrodeposition of magnetic atoms into predefined templates. They were polymer-particle composites made from the assembly of millions of sub-10 nm iron oxide nanoparticles, that were glued together by cationic polymers. The magnetic properties of the wires were inherited from the iron oxide particles and allow to rotate the wires in a propeller-like motion by the application of an external field. [21, 22]

Our approach with cell culture consisted to show that *i)* the wires were actually internalized, and that *ii)* they were biocompatible as the fibroblasts remain viable relative to controls. Concerning the first point, quantitative measurements of internalized amounts revealed that 4 µm nanowires were uptaken in large proportion, typically 35% of the initial seeded quantity. Direct visualization of the wires inside the cytoplasm by optical and electron microscopy also confirmed these conclusions: a large number of incubated wires were able to cross the cell plasma membranes. The results obtained by TEM and immunofluorescence using LAMP1 markers provide a consistent description of the internalized nanowires and do not preclude the possibility of multiple entry mechanisms, including macropinocytosis[13, 16, 17] or perforation of the outer plasma membrane.[5, 20] More experiments are necessary to reach a definite conclusion concerning the portals of entry of the wires into the cells.

Regarding cytotoxicity, MTT assays, cell proliferation and oxidative stress measurements on treated fibroblasts revealed normal responses, even at doses as high as 250 wires per cell ([Fe] = 1 mM). The findings concerning the uptake and the toxicity are in good agreement with recent results on electrodeposited nanowires,[4-6, 15] and on certain types of carbon nanotube samples.[31, 35]

The surprising result revealed in this work concerns the fate of the wires once taken up by the cells. As shown by TEM after a 24 h incubation, the cells were able to degrade the nanowires and to cut them into smaller aggregates, with typical size 200 nm. Interestingly, the wires and the remainders of the degradation were found either in vesicular compartments or directly dispersed in the cytosol. This was not the case for the single nanoparticles which were found in endosomal compartments only. These results suggest that the degradation is likely to occur as a consequence of the internal structure of the wires which is that of a composite material characterized by non-covalent (electrostatic) forces. We anticipate that in the long-term, all the wires should be transformed, avoiding a potential asbestos-like toxicity effect related to high aspect ratio morphologies.[35] From these investigations, it is concluded that the iron oxide based nanowires can be used safely with living cells, *e.g.* as microtools for *in vitro* and *in vivo* applications.

## Materials and Methods

Chemicals, synthesis and characterization

*Nanoparticles synthesis* : The synthesis of iron oxide nanoparticles ($\gamma$-$Fe_2O_3$, maghemite) was based on the polycondensation of metallic salts in alkaline aqueous media elaborated by R. Massart.[36] An image of transmission electron microscopy (TEM) obtained from a dilute aqueous dispersion exhibits compact and spherical particles (Fig. 1a). The size distribution of these particles was described by a log-normal function with a median diameter $D_{NP}$ = 6.7 nm and polydispersity $s_{NP}$ = 0.2 (Fig. 1b). The polydispersity was defined as the ratio between standard deviation and average diameter. Extensive characterization of the particles including the determination of the surface charge, structural anisotropy and magnetization can be found in our previous work.[29, 37] The release amounts and rates of ferric ions for the present dispersions were also estimated in acidic and neutral pH conditions and they were found to be insignificant (SI-8). To improve their colloidal stability, the cationic particles were coated with $M_W$ = 2000 g mol$^{-1}$ poly(acrylic acid) using the *precipitation-redispersion* process.[38] This process resulted in the adsorption of a highly resilient 3 nm polymer layer surrounding the particles. The wires were fabricated using these coated particles, noted as $PAA_{2K}$–$\gamma$-$Fe_2O_3$ in the following.

*Magnetic Nanowires synthesis* : The wire formation resulted from the electrostatic complexation between oppositely charged nanoparticles and copolymers. The copolymer used was poly(trimethylammonium ethylacrylate)-*b*-poly(acrylamide) with molecular weights 11 000 g mol$^{-1}$ for the charged block and 30 000 g mol$^{-1}$ for the neutral block, abbreviated PTEA$_{11K}$-*b*-PAM$_{30K}$ in the sequel of the paper.[39] The protocols applied here consisted first in the screening of the electrostatic interactions by bringing the polymer and particle dispersions to high salt concentration, and second in progressively removing the salt by *dialysis* or by *dilution*. With this technique, PAA$_{2K}$−γ-Fe$_2$O$_3$ and PTEA$_{11K}$-*b*-PAM$_{30K}$ were intimately mixed in solution before they could interact. With decreasing ionic strength, an abrupt transition between a disperse and an aggregated state of particles occurred. The particles and polymers were permanently co-assembled in the presence of a magnetic field to stimulate unidirectional growth. Figs. 1c and 1d display transmission optical and electron microscopy images of the wires. Their average diameters were found to be 200 nm, with lengths ranging from 1 to 40 µm. As an illustration of their magnetic properties, a movie of nanowires subjected to a rotating magnetic field at the frequency of 0.2 Hz is shown in Supporting Information (Movie#1). The wires were actually polydisperse and their length distribution was described by a log-normal function with median value L$_{wire}$ and polydispersity s. For this work, two batches of nanowires were synthesized, one with L$_{wire}$ = 4.1 µm and s = 0.45 and the second with L$_{wire}$ = 15.6 µm and s = 0.65 (Table I). Electrophoretic mobility and ζ–potential measurements using Zetasizer Nano ZS Malvern Instrument show that the wires were electrically neutral.

## Experimental Methods

*Transmission optical Microscopy* : For optical microscopy observations, phase-contrast images of the cells containing wires were acquired on an IX71 inverted microscope (Olympus) equipped with 40× and 60× objectives. 2×10$^4$ NIH/3T3 fibroblast cells were first seeded onto a 96-well plate for 24 h prior incubation with nanowires. µl-aliquots containing nanowires were added to the supernatant. The nanowire concentration was defined by the ratio of the number of wires incubated per cell, here fixed at 30. The incubation of the wires lasted 24 more hours. The third day, excess medium was removed and the cells were washed with PBS solution (with calcium and magnesium, Dulbecco's, PAA Laboratories), trypsinized and centrifuged. Cell pellets were resuspended in Dulbecco's modified Eagle's medium (DMEM). For optical microscopy, 20 µl of the previous cell suspension were deposited on a glass plate and sealed into to a Gene Frame® (Abgene/Advanced Biotech) dual adhesive system. The sample was then left for 4 h in the incubator to let cells adhere onto the glass plate. Images were observed using a Photometrics Cascade camera (Roper Scientific) and Metaview software (Universal Imaging Inc.) as acquisition system. In order to determine the length distribution of the wires, pictures were digitized and treated by the ImageJ software (http://rsbweb.nih.gov/ij/).

*Transmission Electron Microscopy* : TEM on nanomaterials was carried out on a Jeol-100 CX microscope at the SIARE facility of Université Pierre et Marie Curie (Paris 6). It was utilized to characterize the $PAA_{2K}$–γ-$Fe_2O_3$ particles and the $PAA_{2K}$–γ-$Fe_2O_3$/$PTEA_{11K}$-*b*-$PAM_{30K}$ nanowires using magnifications ranging from 10000× to 160000× (Fig. 1). For the TEM studies of cells, the following protocol was applied. NIH/3T3 fibroblast cells were seeded onto the 6-well plate, after the 24 h incubation with 15 µm nanowires, excess medium was removed, and the cells were washed in 0.2 M phosphate buffer (PBS), pH 7.4 and fixed in 2% glutaraldehyde-phosphate buffer 0.1 M for 1 h at room temperature. Fixed cells were washed in 0.2 M PBS. Then, they were postfixed in 1% osmium-phosphate buffer 0.1 M for 45 min at room temperature in dark conditions. After 0.1 M PBS washes, the samples were dehydrated in increasing concentrations of ethanol. Samples were then infiltrated in 1:1 ethanol:epon resin for 1 h and finally in 100% epon resin for 48 h at 60°C for polymerization. 90 nm-thick sections were cut with an ultramicrotome (LEICA, Ultracut UCT) and picked up on copper-rhodium grids. They were then stained for 7 min in 2% uranyl acetate and for 7 min in 0.2% lead citrate. Grids were analyzed with a transmission electron microscope (ZEISS, EM 912 OMEGA) equipped with a $LaB_6$ filament, at 80 kV and images were captured with a digital camera (SS-CCD, Proscan 1024×1024), and the iTEM software.

*MILC protocol* : UV-visible spectrometry was performed in the MILC protocol (**M**ass of metal **I**nternalized/Adsorbed by **L**iving **C**ells) which consists in the measurement of the mass of nanoparticles incorporated into living cells. The quantity to be determined is the mass of iron expressed in the unit of picogram per cell. Cells were seeded onto 3.6 cm Petri dishes, incubated until reaching 60% confluence and then incubated with nanomaterials at different concentrations for 24 h. The concentration ranges explored were [Fe] = 10 µM – 50 mM for the particles and [Fe] = 10 µM – 0.5 mM for the 4 µm nanowires (corresponding to 2 - 120 wires per cell). After the incubation period, the supernatant was removed and the layer of cells washed thoroughly with PBS. The cells were then trypsinized and mixed again with white DMEM without serum. 20 µl-aliquots were taken up for counting using a Malassez counting chamber. The cells were finally centrifuged and pellets were dissolved in 35 vol. % HCl. The cells dissolved in HCl were poured in a UV-Vis microcell, studied with a Variant spectrophotometer (Cary 50 Scan) and calibrated with the help of a reference.[40] A complete description of the MILC protocol is provided in the Supporting Information (SI-2).

*Cell culture and cellular growth* : NIH/3T3 fibroblast cells from mice were grown in T25-flasks as a monolayer in DMEM with high glucose (4.5 g $L^{-1}$) and stable glutamine (PAA Laboratories GmbH, Austria). This medium was supplemented with 10% fetal bovine serum (FBS) and 1% penicillin/streptomycin (PAA Laboratories GmbH, Austria), referred to as cell culture medium. Exponentially growing cultures were maintained in a humidified atmosphere of 5% $CO_2$ and 95% air at 37°C, and in these conditions the plating efficiency was 70 – 90% and the cell duplication time was 12 – 14 h. Cell cultures were passaged twice weekly using trypsin–EDTA (PAA Laboratories GmbH, Austria) to detach the cells from their culture flasks and wells. The cells were pelleted by centrifugation at 1200 rpm for 5 min.

Supernatants were removed and cell pellets were re-suspended in assay medium and counted using a Malassez counting chamber. Cellular growth was measured with both untreated cells and cells treated for 24 h with different concentrations of 4 µm nanowires ranging from 3 to 60 wires per cell and counted in a Malassez chamber.

*MTT toxicity assays*: MTT assays were performed with $PAA_{2K}$-coated iron oxide nanoparticles at [Fe] = 10 µM - 10 mM and with 4 µm nanowires at [Fe] = 0.3 µM - 1 mM, corresponding to 0.07 – 250 wires per cell. These concentration domains are within the ranges reported in the literature for *in vivo*[41] and *in vitro*[24, 26] assays. Cells were seeded into 96-well microplates, and the plates were placed in an incubator overnight to allow attachment and recovery. Cell densities were adjusted to $2\times10^4$ cells per well (200 µl). After 24 h, the nanoparticles and nanowires were applied directly to each well using a multichannel pipette. Cultures were incubated in triplicate for 24 h at 37°C. The MTT assay depends on the cellular reduction of MTT (3- (4,5-dimethylthiazol-2-yl)-2,5-diphenyl tetrazolium bromide, Sigma Aldrich Chemical) by the mitochondrial dehydrogenase of viable cells forming a blue formazan product which can be measured spectrophotometrically.[42] MTT was prepared at 5 mg ml$^{-1}$ in PBS (with calcium and magnesium, Dulbecco's, PAA Laboratories) and then diluted 1:5 in medium without serum and without Phenol Red. After 24 h of incubation with nanoparticles, the medium was removed, the wells were washed twice with 100µl PBS (1X) and 200 µl of the MTT solution was added to the microculture wells. After 4 h incubation at 37°C, the MTT solution was removed and 100 µl of 100% DMSO was added to each well to solubilize the MTT–formazan product. The microplate was then placed on a ferrite magnet to separate the magnetic material from the supernatant, and 50µl of this solubilized formazan was transferred to another plate for further treatment. The absorbance at 562 nm was then measured with a microplate reader (Perkin Elmer) and the results were expressed as the percentage of control cells. A series of controls performed to assess the potential interactions between the nanomaterials with the MTT and with the formazan crystals were conducted and exhibited negative results, as shown in SI-9. Additional controls and references were also monitored without particles on populations ranging from $5\times10^3$ to $5\times10^5$ cells. In our studies, the viability remained around 100% within the experimental accuracy. Values above the 100 % limit (e.g. 116 % at 0.1 mM for the $PAA_{2K}$–$\gamma$-$Fe_2O_3$ particles, Fig. 6a) were ascribed to an uneven seeding of the cells onto the 96-well plate rather than to an actual increase in cell proliferation. Finally, it was checked that the formazan absorption was not contaminated by the presence of the nanomaterials, and that this assay reflects actual mitochondrial activity.

*Oxidative stress (DHE)*: Generation of reactive oxygen species (ROS) was evaluated by dihydroethidium (DHE) probe (SIGMA) known to be oxidized by intracellular superoxide anions. Cells were treated for 4 h with different concentrations of nanowires ([Fe] = 0.1 mM, 0.3 mM and 0.7 mM, corresponding to 20, 70 and 170 wires/cell) or with various doses of $H_2O_2$ as positive control. After trypsinization cells were centrifuged at 1200 rpm for 5 min and resuspended in cell culture medium containing 1µM DHE. The analysis of the bright red fluorescence of the oxidized DNA intercalated probe was performed with a CyAn ADP cytometer.[32]

*Immunofluorescence and microscopy :* NIH/3T3 cells were fixed in 4% PFA-PBS for 45 min at 4°C, incubated for 10 min with 50 mM $NH_4Cl$-PBS, washed twice in 2 % FCS-PBS (PBS-FCS) and incubated for 45 min with the anti-mouse Lamp1 (CD107a) antibodies (rat monoclonal, clone 1D4B, Becton Dickinson) in the permeabilizing buffer (PBS-FCS/0.05 % saponin) to detect late endosomal compartments. Subsequent steps were performed at room temperature in permeabilizing buffer. After 2 washes, cells were incubated with Cy3-labeled F(ab')$_2$ anti-rat IgG antibodies (Jackson Immunoresearch) in permeabilizing buffer, washed 3 times in the same buffer and twice in PBS, and mounted on microscope slides in 100 mg/ml Mowiol, 25 % (v/v) glycerol, 100 mM Tris, pH 8. Samples were examined under an inverted wide-field microscope (Leica DMI6000) equipped with an oil immersion objective (100x PL APO HCX, 1.4 NA) and a cooled CCD camera (MicroMAX 1300Y/HS, Princeton Instruments). Z-stacks of wide-field fluorescent images were acquired using a piezo at 0.2 μm increments. The 3D recontructions of the endosomes containing nanowires (SI-7) were made using the IMARIS software (BITPLANE Scientific Software).

# Acknowledgments

We thank Loudjy Chevry, Jérémie Courtois, François Darchen, Claire Desnos, Jérôme Fresnais, Jean-Pierre Henry, Olivier Sandre and Michel Seigneuret for fruitful discussions. The Laboratoire Physico-chimie des Electrolytes, Colloïdes et Sciences Analytiques (UMR Université Pierre et Marie Curie-CNRS n° 7612) is acknowledged for providing us with the magnetic nanoparticles. This research was supported in part by Rhodia (France), by the Agence Nationale de la Recherche under the contracts BLAN07-3_206866 and ANR-09-NANO-P200-36, by the European Community through the project : "*NANO3T—Biofunctionalized Metal and Magnetic Nanoparticles for Targeted Tumor Therapy*", project number 214137 (FP7-NMP-2007-SMALL-1) and by the Région Ile-de-France in the DIM framework related to Health, Environnement and Toxicology (SEnT). The oxidative stress experiments and analysis were performed with a CyAn-ADP cytometer (contract number: R03/75-79) at the Institut Jacques Monod Paris-Diderot and financed by Ligue Nationale contre le Cancer (Comité Ile–de-France). We also thank the *Service of Electron Microscopy* IFR83 for providing TEM installation and the *Cochin Imaging Facility* for the image deconvolution of the LAMP1 images.

# Supporting Information

The Supporting Information section shows how to analyze the Brownian motions of internalized wires in terms of mean square angular displacement and rotational diffusion constant (SI-1). Details of the protocol for measuring the mass of metal internalized/adsorbed by living cells (MILC) are provided in SI-2. Proliferation assays obtained for the fibroblasts incubated with $PAA_{2K}$–γ-$Fe_2O_3$ particles (SI-3) are shown to allow the comparaison with the data of Fig. 5. Controls of the oxidative properties of the particles/wires (SI-4) and on the effect of the pH on the nanowire stability (SI-5) are included. Additional TEM materials are displayed in SI-6 to prove that isolated nanoparticles or small clusters of particles can be found in the cytosol. 3D reconstruction of the of the immunofluorescence images are illustrated in SI-7. The release amounts and release rates of ferric ion Fe3+ at neutral and acidic pH for the Massart dispersions used in this work are estimated in SI-8. Additional control experiments show the absence of interaction between the MTT/formazan crystals with the nanomaterials (SI-9) and finally a statistical study of the iron oxide loaded compartments for cells incubated with PAA$_{2K}$–γ-Fe$_2$O$_3$ particles is provided in SI-10. Four movies complete this section. In Movie#1, a nanowire is subjected to a rotating magnetic field (B = 0.01 T) at the frequency of 0.2 Hz and in Movie#2 the Brownian motions of magnetic nanowires inside NIH/3T3 cells are observed. Movie#3 and #4 show Z-stacks of images taken every 0.2 µm for fixed NIH/3T3 cells, in phase contrast and fluorescence respectively. This information is available free of charge *via* the Internet at http://pubs.acs.org/.